\documentclass{aa}
\usepackage[varg]{txfonts}
\usepackage[utf8]{inputenc}
\usepackage{siunitx}

\usepackage[colorlinks=true, linkcolor=blue, citecolor=blue, urlcolor=blue]{hyperref}
\usepackage{natbib}

\renewcommand\makeLineNumber{}

\bibpunct{(}{)}{;}{a}{}{,}

\newcommand{\p}[2]{\dfrac{\partial #1}{\partial #2}}

\newcommand{\paren}[1]{\Biggl( #1 \Biggr)}
\begin{document}

\title{A kinetic model of solar wind acceleration driven by ambipolar electric potential and velocity-space diffusion}
\author{M. P\'eters de Bonhome\thanks{e-mail: maximilien.petersdebonhome@kuleuven.be}\inst{\ref{inst1},\ref{inst2}}
\and V. Pierrard\inst{\ref{inst2},\ref{inst3}}
\and F. Bacchini\inst{\ref{inst1},\ref{inst2}}}

\institute{Centre for mathematical Plasma Astrophysics, Department of Mathematics, KU Leuven, Celestijnenlaan 200B, B-3001 Leuven, Belgium\label{inst1}
\and Space Physics and Solar-Terrestrial Centre of Excellence, Royal Belgian Institute for Space Aeronomy, Avenue Circulaire 3, B-1180 Brussels, Belgium\label{inst2}
\and Earth and Life Institute Climate Sciences ELI-C, Université Catholique de Louvain, Place Louis Pasteur 3, B-1348 Louvain-la-Neuve, Belgium\label{inst3}}

\date{Received 24 February 2025 / Accepted 29 March 2025}

\abstract {Parker Solar Probe (PSP) observations have revealed that most of the solar wind acceleration occurs very close to the Sun. This acceleration is partly due to the global electric potential originating from the mass disparity between electrons and protons, coupled with the constraints of charge quasi-neutrality and zero-current conditions in the solar wind plasma. However, the exact mechanism that accounts for the remaining acceleration has not yet been identified.} {We aim to provide a framework that incorporates the electric-field-driven component of the acceleration while also introducing an additional acceleration mechanism via a velocity-space diffusion of the particles. This will help us determine the extent of extra acceleration, beyond the electric-field-driven component, required to fully reproduce the acceleration of the solar wind in theoretical models.} {We modified an existing kinetic exospheric model to account for the unexplained solar wind acceleration by including velocity-space diffusion, thereby capturing the effect of collisions and wave-particle interactions within the exospheric approach. We compared the electric field derived from the sunward deficit of velocity distribution functions observed by PSP between 13.3 and 50 solar radii ($R_s$) with the electric field found self-consistently by the kinetic exospheric model.} {The effect of velocity-space diffusion is found to reduce the temperature anisotropy and impact the solar wind acceleration while leaving the electric potential unchanged. The approach described in this work enables  the diffusion to be adjusted to effectively reduce or increase the solar wind acceleration. Even without diffusion, the model is able to reproduce the anticorrelation between the electric potential and the solar wind terminal velocity found by PSP. This suggests that the electric potential might still be of major importance in explaining the solar wind acceleration.} {}


\titlerunning{A kinetic model of the solar wind acceleration}
\authorrunning{P\'eters de Bonhome et al.}

\maketitle

\section{Introduction}

The electrons and protons that are blown away from the Sun through interplanetary space are the main constituents of a continuous outflow of particles referred to as the solar wind. The exact mechanism that allows the solar wind to reach observed velocities averaging between 300 and 800~$\si{km.s^{-1}}$ remains one of the most significant open questions in space physics \citep[e.g.,][]{raouafi_solar_2021}. Since the first in situ solar wind observations in 1959, neither magnetohydrodynamic nor kinetic models have been able to self-consistently explain the solar wind acceleration. In particular, magnetohydrodynamic models are not able to capture the origin of the fast solar wind with realistic coronal temperatures and often require additional heating mechanisms (\citealt{parker_dynamical_1965,leer_acceleration_1982,hansteen_solar_2012}). Although some kinetic approaches can provide an explanation for the presence of peculiar observed velocity distribution functions (VDFs) of solar wind particles and propose potential mechanisms of solar wind acceleration (e.g., \citealt{hollweg_generation_2002,vocks_electron_2005,seough_strahl_2015}), they generally lack any global electric-field-driven component of the acceleration. We investigated this acceleration by improving a specific class of kinetic model known as exospheric models, which are fundamentally based on considering the electric-field-driven component as the primary acceleration mechanism. A comprehensive review of these models has been provided by \citet{fahr_modern_1983}.

The first exospheric models were developed for planetary exospheres and were later adapted for the solar wind (\citealt{chamberlain_interplanetary_1960}). These models describe the radial evolution of the particle VDFs by neglecting collisions between particles. The first model of the solar wind that included a consistent ``ambipolar'' electric field was developed by \cite{lemaire_kinetic_1971} for a collisionless regime above the solar surface at a given altitude called the ``exobase,'' which was chosen to be at 6 solar radii ($R_s$). This electric field is considered to be ambipolar in the sense that it satisfies the quasi-neutrality and zero-current conditions between protons and electrons observed to be preserved in the solar wind \citep[e.g.,][]{feldman_solar_1975}. The challenging aspect of these models was to determine the radial evolution of this ambipolar electric field self-consistently with the exospheric approach, especially in the case of non-monotonic proton potential (typically found when assuming an exobase below 6~$R_s$, which better describes the solar wind case), as noted by \cite{jockers_solar_1970}. \cite{lamy_kinetic_2003}, inspired by \cite{jockers_solar_1970}, developed a generalized method for approximating the solution for this self-consistent electric potential.

The use of Kappa (also known as Lorentzian) VDFs for electrons, in contrast to Maxwellian VDFs for protons, constitutes one of the major breakthroughs that was first introduced in an exospheric model by \cite{pierrard_lorentzian_1996} and applied to the solar wind by \cite{maksimovic_kinetic_1997}. This approach was motivated by the need to account for the so-called suprathermal electrons observed extensively in the solar wind (e.g., \citealt{lin1980energetic,pilipp_characteristics_1987,pierrard_halo2016,abraham_radial_2022,salem_precision_2023}). Such suprathermal electron populations are better described by a Kappa distribution function rather than a single Maxwellian distribution, as shown, for instance, by \cite{maksimovic_ulysses_1997} and more recently by \cite{Zheng_kappa2024}. These electrons are high-energy particles that do not follow the classical single Maxwellian distribution fitted to the lower-energy part (called the core) of the observed distribution. These suprathermal distributions are typical for the halo, which extends in all directions, and the electron strahl, a more concentrated component observed in the anti-sunward direction (e.g.,  \citealt{feldman_solar_1975,pilipp_characteristics_1987,pierrard_strahl2001}). The mathematical description of Kappa distributions required for exospheric models was first introduced by \cite{pierrard_lorentzian_1996} and generalized for non-monotonic potentials by \cite{lamy_kinetic_2003}. Within the scope of this model, using a Kappa distribution for electrons increases the number of electrons with high speeds in the anti-sunward direction, which effectively increases the electric potential difference between any two radial distances (see \citealt{pierrard_kappa_2010} for a review). Therefore, this allows more protons to reach the maximum of the non-monotonic proton potential (usually within 1 to 5~$R_s$), which in turn accelerates the wind to realistically high bulk speeds without requiring unrealistically high temperatures. Following the same approach, \cite{zouganelis_transonic_2004} showed that similar results could also be obtained with a sum of two Maxwellians or a sum of a Maxwellian and a Kappa distribution (for the core and strahl), which would not limit the significance of the strahl to the power law factor, $\kappa$.

Given the simplicity and effectiveness of the \cite{lamy_kinetic_2003} approach in assigning at least part of the unexplained acceleration to the presence of suprathermal electrons, the model was later extended to minor ions (ions with lower abundances than protons) by \cite{pierrard_coronal_2014}. More recently, an improved version of this exospheric model, incorporating regularized Kappa distribution functions (which allow even lower $\kappa$ values and therefore a higher fraction of suprathermal particles), was compared with Parker Solar Probe (PSP) and Solar Orbiter observations by \cite{pierrard_exospheric_2023}. Their results demonstrate that this model reproduces the radial evolution of the solar wind bulk velocity but cannot accurately reproduce the observed anisotropy in the solar wind. This discrepancy arises from the intrinsic conservation of energy and magnetic moment, which strictly constrain modeled VDFs in velocity space.

It has already been noted on multiple occasions that the idealized view provided by the exospheric model is not completely accurate. In particular, collisions and plasma waves are expected to be the mechanisms responsible for the scattering of electrons in the solar wind (\citealt{pierrard_selfconsistent_2001,salem_electron_2003,pierrard_evolution_2011,boldyrev_electron_2020,micera_particle--cell_2020,verscharen_electron-driven_2022,yoon_regulation_2024}). These effects inevitably populate regions of velocity space that would otherwise be inaccessible and are believed to be the main mechanisms responsible for the presence of the so-called halo population of higher-energy electrons that are not preferentially anti-sunward directed, unlike the strahl (\citealt{feldman_solar_1975,pilipp_characteristics_1987,maksimovic_ulysses_1997,stverak_electron_2008,stverak_radial_2009}). Therefore, we propose a way to simulate the effect of particle scattering on the moments of the electron VDF that are modeled by introducing a new, ``diffused'' population of particles.

We aim to improve the existing kinetic exospheric model by first properly describing the exospheric framework applied to the solar wind. Therefore, Sect.~\ref{seq:exo} provides a review of an existing kinetic exospheric model. Section~\ref{seq:diff} describes how we incorporated the velocity-space diffusion within the reviewed kinetic exospheric approach. In particular, we investigate how diffused electrons affect the solar wind temperature, bulk velocity, and density in Sect.~\ref{seq:res}. The relevance of the kinetic exospheric model is addressed in Sect.~\ref{seq:dis}, as it effectively explains one of the components responsible for solar wind acceleration through its self-consistent electric field, derived from the quasi-neutrality and zero-current conditions. Finally, the conclusions and future perspectives are provided in Sect.~\ref{seq:concl}.

\section{Review of the solar wind kinetic exospheric model}

\label{seq:exo}

The kinetic exospheric model is based on the assumption that the solar wind becomes collisionless above a radial distance $r_0$, called the exobase. This is generally defined by considering that the ratio of the plasma density scale height to the particle mean free path (called the Knudsen number), must be equal to one for this assumption to be valid. More precisely, protons and electrons in the solar wind have different reference altitudes at which their respective Knudsen number reaches unity, as electrons typically have a shorter mean free path than protons (\citealt{maksimovic_kinetic_1997}). However, for simplicity, we did not make this distinction and instead considered a simple, sharp transition between collision-dominated and collisionless regions at $r_0 = 1.2~R_s$. This exobase level was arbitrarily chosen to emphasize the model's physics while remaining generally close to the expected position of the actual exobase above the solar surface (\citealt{pierrard_exospheric_2023}). We began our mathematical formulation of the model from the general kinetic equation. 

In kinetic models, the evolution of the VDF $f_i(\vec{r},\vec{v},t)$ for the particle species $i$, as a function of position $\vec{r}$, velocity $\vec{v}$, and time $t$, is given by the Liouville equation,
\begin{equation}
    \p{f_i}{t} + \vec{v}\cdot\p{f_i}{\vec{r}} + \frac{\vec{F}}{m_i}\cdot\p{f_i}{\vec{v}} = \left(\frac{\text{d}f_i}{\text{d}t}\right)_c.
    \label{eqLiou}
\end{equation}
In this equation, $m_i$ is the mass of the particle considered and 
\begin{equation}
    \vec{F} = m_i \vec{g} + q_i\vec{E} + q_i(\vec{v} \times \vec{B})
\end{equation}
is the force applied to each particle with $\vec{g}$ being the gravitational acceleration, $q_i$ the charge of the particle, $\vec{E}$ the electric field, and $\vec{B}$ the magnetic field. The right-hand side of Eq.~\eqref{eqLiou} corresponds to particle--particle interactions. Considering the collisionless assumption in the exosphere (above the exobase), we can reduce this equation to the Vlasov equation,
\begin{equation}
    \p{f_i}{t} + \vec{v}\cdot\p{f_i}{\vec{r}} + \frac{\vec{F}}{m_i}\cdot\p{f_i}{\vec{v}} =   0.
    \label{eqLiou2}
\end{equation}
The Liouville theorem asserts that the VDF is constant along the trajectories of the system. This means that any function $f(c_1, c_2, ..., c_n)$, where $c_1, c_2, ..., c_n$ are constants of the motion through external forces $\vec{F}$, is a solution to Eq.~\eqref{eqLiou2}. In the case of the solar wind, the constants of the motion for each particle are the total (kinetic plus potential) energy,
\begin{equation}
    \mathcal{E} = \frac{m_i v^2}{2} + m_i \phi + Z_ieV = c_1,
\end{equation}
and the magnetic moment,
\begin{equation}
    \mathcal{M} = \frac{m_iv^2_\perp}{2B} = c_2,
\end{equation}
where $\phi =-GM_s/r$ is the gravitational potential, $G$ is the gravitational constant, $M_s$ is the mass of the Sun, $Z_i$ is the atomic number of the particle species $i$, $e$ is the fundamental charge, $V$ is the electric potential, and $v_\perp$ is the speed perpendicular to the magnetic field direction. Therefore, if we assume a VDF that only depends on $\mathcal{E}$ and $\mathcal{M}$ at $r_0$, we can deduce its evolution throughout the whole radial expansion. In our model, we assume a Maxwellian VDF for protons at any $r \geq r_0$ given by
\begin{equation}
    f_p(r,\mathcal{E}_p) = N_p\left(\frac{1}{\pi w_p^2}\right)^{3/2}\exp\left(-\frac{\mathcal{E}_p}{k_b T_{0p}}\right),
    \label{eq:fp}
\end{equation}
where $N_p$ is a factor of normalization that allows us to retrieve the specified proton density at $r_0$ by integrating the proton VDF in velocity space, $w_p=\sqrt{2 k_b T_{0p}/m_p}$ is the proton thermal velocity with $k_b$ being the Boltzmann constant, $T_{0p}$ being the proton temperature at $r_0$, $m_p$ being the proton mass, and $\mathcal{E}_p/(k_b T_{0p})$ is the normalized proton energy such that
\begin{equation}
    \frac{\mathcal{E}_p}{k_b T_{0p}} \equiv \frac{m_p v^2}{2 k_b T_{0p}} + \frac{m_p \Delta\phi + Z_p e \Delta V}{k_b T_{0p}} = \frac{v_{\parallel p}^2+v_{\perp p}^2}{w_p^2} + \psi_p(r),
    \label{adiEp}
\end{equation}
with $\Delta\phi \equiv \phi(r) - \phi(r_0)$, $\Delta V \equiv V(r) - V(r_0)$, $v_{\parallel p}$ and $v_{\perp p}$ being the parallel and perpendicular speed of protons with respect to the direction of the magnetic field $\vec{B}$, and 
\begin{equation}
    \psi_p(r) = \frac{2}{w_p^2} \left(\phi(r) + \frac{Z_p e V(r)}{m_p} - \phi(r_0) - \frac{Z_p e V(r_0)}{m_p}\right)
\end{equation}
being the normalized total potential for protons. For electrons, the assumed VDF at $r$ is a Kappa distribution given by 
\begin{equation}
    f_e(r,\mathcal{E}_e) = 
    N_e\left(\frac{1}{\pi\kappa w_e^2}\right)^{3/2}\frac{\Gamma(\kappa+1)}{\Gamma(\kappa-1/2)} 
    \left(1+\frac{\mathcal{E}_e}{(\kappa-3/2) k_b T_{0e}}\right)^{-\kappa-1}.
    \label{eq:fe}
\end{equation}
Here, $\kappa$ is the parameter associated with a power law decreasing Kappa VDF, instead of an exponentially decreasing Maxwellian VDF (i.e., for $\kappa \rightarrow \infty$ the Maxwellian distribution is retrieved); $w_e=\sqrt{(2\kappa-3) k_b T_{0e}/(\kappa m_e)}$ is the electron thermal velocity, $T_{0e}$ is the electron temperature at $r_0$, $m_e$ is the electron mass, $\Gamma$ is the Gamma function, $N_e$ is a factor of normalization (defined in Sect. \ref{seq:diff}) that allows us to retrieve the specified electron density at $r_0$ by integrating the electron VDF in velocity space, and $\mathcal{E}_e/[(\kappa-3/2) k_b T_{0e}]$ is the normalized electron energy such that
\begin{equation}
    \begin{split}
    \frac{\mathcal{E}_e}{(\kappa-3/2) k_b T_{0e}} & \equiv  \frac{m_e v^2}{2 (\kappa-3/2) k_b T_{0e}} + \frac{m_e \Delta\phi + Z_e e \Delta V}{(\kappa-3/2) k_b T_{0e}} \\
    & \equiv \frac{(v_{\parallel e}^2+v_{\perp e}^2)}{\kappa w_e^2}+\frac{\psi_e(r)}{\kappa},
    \end{split}
    \label{adiEe}
\end{equation}
with $v_{\parallel e}$ and $v_{\perp e}$ being the parallel and perpendicular speed of electrons with respect to the direction of $\vec{B}$, and
\begin{equation}
    \psi_e(r) = \frac{2}{w_e^2} \left(\phi(r) + \frac{Z_e e V(r)}{m_e} - \phi(r_0) - \frac{Z_e e V(r_0)}{m_e}\right)  
\end{equation}
being the normalized total potential for electrons. We note that $f_p$ and $f_e$ are functions of $\mathcal{E}$ but also $\mathcal{M}$ since the magnetic moment will have an influence on the allowed perpendicular velocities of particles depending on their origin. 

Considering that $\mathcal{E}$ and $\mathcal{M}$ are constants for each particle, we can write
\begin{equation}
    v_{\perp i}^2(r_0) + v_{\parallel i}^2(r_0) + w_i^2 \psi_i(r_0) = v_{\perp i}^2(r) + v_{\parallel i}^2(r) + w_i^2 \psi_i(r)
    \label{constE}
\end{equation}
and
\begin{equation}
    \frac{v_{\perp i}^2(r_0)}{B(r_0)} = \frac{v_{\perp i}^2(r)}{B(r)}.
    \label{constM}
\end{equation}
These equations are given in a general form for electrons and protons, as noted by the species index $i$, which will be omitted, for convenience, in the remainder of the description. If we assume that the reference altitude, $r_0$, is exactly the point where the solar wind becomes collisionless and that below $r_0$ the particles are dominated by collisions (in such a way that there is no preferential direction for the particle motion), then at $r_0$ all particles should be constrained to $v_\parallel(r_0) \geq 0$. Now, if we consider the lower limit $v_\parallel(r_0) = 0$, and we combine Eqs.~\eqref{constE} and~\eqref{constM}, we get  
\begin{equation}
    v_\perp^2(r)\left(\frac{B(r_0)}{B(r)} - 1\right) =  v_\parallel^2(r) + w^2 \psi(r),
\end{equation}
with $\psi(r_0) = 0$. In our model, we further assume, for simplicity, that the magnetic field is purely radial and that $B(r) \propto 1/r^2$. This approximation is justified because \cite{pierrard_collisionless_2001} demonstrated that incorporating a spiral magnetic field into the kinetic exospheric model alters the radial dependence of the magnetic field without significantly affecting the resulting acceleration. By introducing the factor $\eta \equiv (r_0/r)^2$, we can therefore rewrite 
\begin{equation}
    v_\perp^2(r) = \left(\frac{\eta}{1-\eta}\right)\left(v_\parallel^2(r) + w^2 \psi(r)\right),
    \label{hyp}
\end{equation}
for the case $v_\parallel(r_0) = 0$. 

\begin{table}[!b]
    \caption{\label{tab:inputs} Input parameters of the kinetic exospheric model used for the VDFs shown in Fig.~\ref{fig:VDFcomparison}.}
    \begin{center}
        \begin{tabular}{c c c c}
        Parameter name & Notation & Value & Unit\\
        \hline
        Exobase level & $r_0$ & 1.2 & $R_s$  \\
        Electron temp. at exobase & $T_{0e}$ & 1.5 & \si{MK} \\
        Proton temp. at exobase & $T_{0p}$ & 1.2 & \si{MK} \\
        Density at exobase & $n_0$ & $1 \times 10^{13}$ & \si{m^{-3}} \\
        Kappa & $\kappa$ & 6 & / \\
        \end{tabular}
    \end{center} 
\end{table}

\begin{figure*}[!t]
    \centering
    \renewcommand{\arraystretch}{0.5}
    \begin{tabular}{cc} %
        \includegraphics{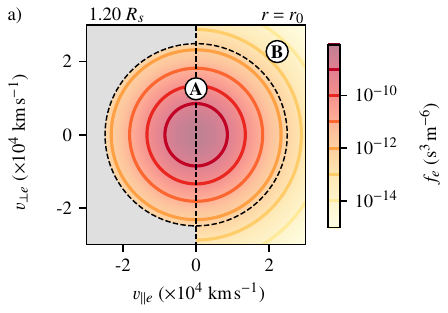} & 
        \includegraphics{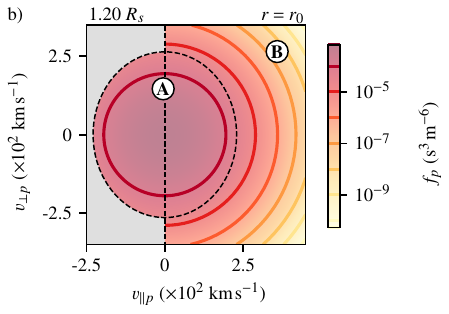} \\ 
        \includegraphics{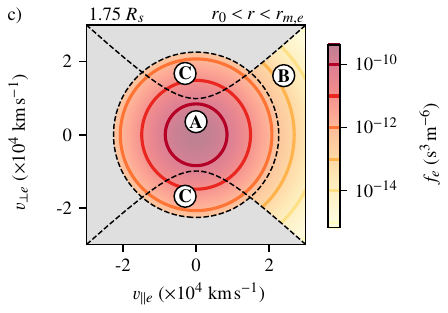} &
        \includegraphics{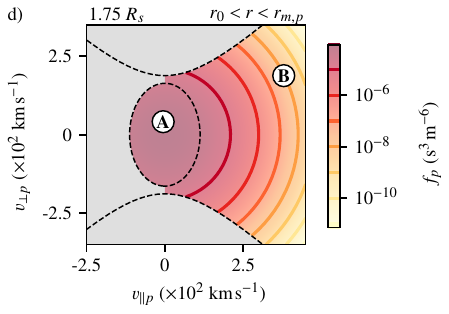} \\ 
        \includegraphics{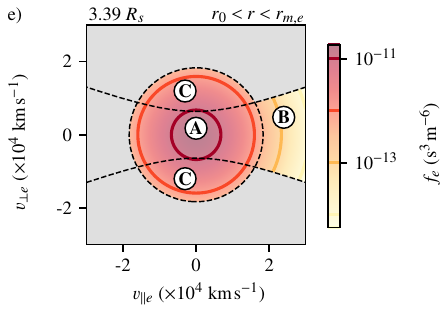} &
        \includegraphics{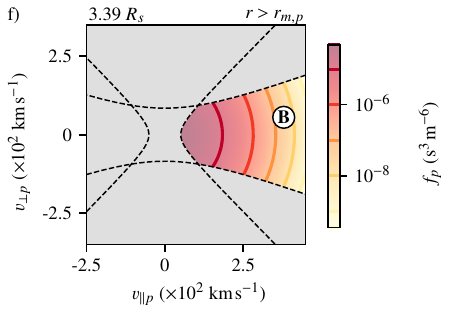}
    \end{tabular}
    \caption{Comparison of the standard exospheric model's computed VDFs for electrons at different radial distances (panels a, c, and e) and for protons at the same radial distances (panels b, d, and f). The $x$-axis is the parallel velocity, and the $y$-axis is the perpendicular velocity with respect to the magnetic field direction. The gray areas correspond to empty regions of velocity space.}
    \label{fig:VDFcomparison}
\end{figure*}

Let us now assume that the potential difference $\psi(r)$ (assuming that $\psi(r_0) = 0$) is increasing at first and then reaches a maximum at $r_{m,i}$ where the index $i$ corresponds to $e$ or $p$ for electrons or protons, respectively (this index will also be omitted). This situation would imply that some particles cannot reach this maximum because they do not possess enough energy. To distinguish between these particles and those that can instead reach this maximum, we defined a velocity threshold characterized by $v_\parallel(r_m) = 0$, since any particle with a positive parallel velocity at $r_m$ will escape from the potential well. By replacing $r_0$ with $r_m$ in Eqs.~\eqref{constE} and~\eqref{constM} and introducing the factor $\mu \equiv (r_m/r)^2$, we similarly obtain 
\begin{equation}
    v_\perp^2(r) = \left(\frac{\mu}{1-\mu}\right)\left(v_\parallel^2(r) + w^2 \left(\psi\left(r\right) - \psi\left(r_m\right)\right)\right),
    \label{ell}
\end{equation}
in the case $v_\parallel(r_m) = 0$. 

These equations were derived by \cite{lemaire_model_1970} for planetary exospheres and are contextualized here in the case of the solar wind in Fig.~\ref{fig:VDFcomparison}. The VDFs of this figure are reconstructed from the model results obtained using Eqs.~\eqref{hyp} and \eqref{ell} and the input parameters of Table \ref{tab:inputs}. Panels a and b show the electron and proton VDFs, respectively, at $r = r_0$. Both species are placed in an increasing potential near the Sun because of the mass difference between protons and electrons. Since the electrons are less impacted by the gravity field than the protons, this ultimately results in generating a charge separation electric field (which accelerates the protons and slows the electrons) to maintain  quasi-neutrality and zero net current between protons and electrons. However, for the protons, the dominant effect is still the gravity field near the Sun, which places them in an attractive potential at first. The boundary at $v_\parallel = 0$ implies that the only particles that can have a $v_\parallel < 0$ at $r = r_0$ are the particles that were reflected due to the potential. This is specifically the case for the particles that could not reach $r_m$ and had to turn back to the Sun. These particles are directly deduced from Eq.~\eqref{ell} with $r_{m,e} \rightarrow \infty$ for electrons and $r_0 \le r_{m,p} \le \infty$ for protons. There are empty regions of velocity space since we did not consider any pitch angle scattering nor the presence of particles coming from infinity. Panels c and d correspond to the radial distance $r_0 < r < r_{m,p}$, while panels e and f correspond to the radial distance $r > r_{m,p}$. In panels c and e, the evolution of Eqs.~\eqref{hyp} and~\eqref{ell} for electrons is represented by dashed lines forming hyperbolas and circles, respectively. The same can be seen for protons in panels d and f, except that here the protons are placed in an attractive potential ($r_0 < r < r_{m,p}$) and then in a repulsive potential ($r > r_{m,p}$). This is why the ellipse dashed line of panel d transforms into a hyperbola with its foci on the $v_\parallel$-axis. 

These dashed lines delimit the regions of velocity space labeled \textbf{A}, \textbf{B}, and \textbf{C} in Fig.~\ref{fig:VDFcomparison}. The \textbf{A} region corresponds to the particles that do not have sufficient energy to overcome the potential maximum; these are called ``ballistic'' particles. The \textbf{B} region corresponds to particles that have enough energy to overcome this potential maximum and are therefore called ``escaping'' particles. The \textbf{C} region corresponds to trapped particles that lack sufficient energy to overcome the potential maximum and cannot reach $r_0$. Instead, they are perpetually reflected because of magnetic moment conservation. In principle, such particles cannot originate directly from $r_0$. However, as generally assumed (e.g., \citealt{maksimovic_kinetic_1997,lamy_kinetic_2003,zouganelis_transonic_2004}), our model still accounts for them because this region is expected to eventually become saturated with particles: any particle entering this region of velocity space (whether through rare collisions or interactions with plasma waves) will remain there indefinitely unless scattered out. Therefore, we consider that the VDF of region \textbf{C} is also described by Eqs.~\eqref{eq:fp} and \eqref{eq:fe} for protons and electrons, respectively, ensuring VDF continuity between ballistic and trapped particles. We note that, in our example, there is no region corresponding to trapped protons (Fig.~\ref{fig:VDFcomparison}d); however, this is not always the case. In fact, with the right parameters, trapped particles could be obtained close to $r_0$ (see, e.g., \citealt{lamy_kinetic_2003}).

To compute the VDFs of our model, we had to determine the electric potential $V(r)$ since both the normalized potentials, $\psi_p(r)$ and $\psi_e(r),$ depend on it. Here, this electric potential is  found self-consistently through the use of the quasi-neutrality between charges and zero-current equations, which are assumed to be strictly respected over sufficiently large scales in the solar wind. To derive $V(r)$, we defined two key quantities. First, the total normalized potential difference of protons between $r_m \equiv r_{m,p}$ and $r_0$ was defined as
\begin{equation}
        \psi_p(r_m) = \frac{m_p \phi_m + eV_m}{k_b T_{0p}} - \frac{m_p \phi_0 + eV_0}{k_b T_{0p}},
\end{equation}
with $\phi_m \equiv \phi(r = r_m)$, $\phi_0 \equiv \phi(r = r_0)$, $V_m \equiv V(r = r_m)$ and $V_0 \equiv V(r = r_0)$. Similarly, the normalized potential difference of electrons between $r_0$ and $r_\infty$ was defined as
\begin{equation}
    \begin{split}
        \psi_e(r_\infty) = \paren{\frac{m_e \phi_\infty - eV_\infty}{k_b T_{0e}} - \frac{m_e \phi_0 - eV_0}{k_b T_{0e}}} \frac{2\kappa}{2\kappa-3},
    \end{split}
\end{equation}
with $\phi_\infty \equiv \phi(r = \infty)$, and $V_\infty \equiv V(r = \infty)$. The model finds the solution of $\psi_p(r_m)$ and $\psi_e(r_\infty)$ through the use of the equality of fluxes (for zero-current) and the equality of densities for protons and electrons (for quasi-neutrality) at $r_m$, which reads 
\begin{equation}
    \begin{split}
            & F_p(r_m) = F_e(r_m), \\
            & n_p(r_m) = n_e(r_m).
    \end{split}
    \label{eq:zc_qn}
\end{equation}
The solutions for $\psi_p(r_m)$, $\psi_e(r_\infty)$, and $r_m$ are determined through an iterative method, employing these equations and the requirement that $\psi_p(r)$ must be unique and continuous (\citealt{jockers_solar_1970}, \citealt{lamy_kinetic_2003}). This enables the calculation of the potentials $\psi_p(r)$ and $\psi_e(r)$, which are subsequently applied to compute macroscopic quantities using moments equations, provided by \cite{lamy_kinetic_2003} and \cite{pierrard_lorentzian_1996}, involving the integrals of the VDFs in different regions of velocity space (i.e., \textbf{A}, \textbf{B}, or \textbf{C}) for protons and electrons, respectively.

\section{Exospheric approach on diffusion}

\label{seq:diff}

The main limitation of ordinary kinetic exospheric models of the solar wind is their inability to self-consistently define a halo population outside the regions of velocity space described in the previous section. To solve this problem, we introduced a halo population that is directly dependent on the number of particles in the classically filled regions of velocity space (\textbf{A}, \textbf{B}, and \textbf{C}). In this way, the radial evolution of the total density stays unchanged while the electrons are redistributed in velocity space so that a tenuous halo appears for greater perpendicular velocities.  

\begin{figure*}[!t]
    \centering
    \includegraphics{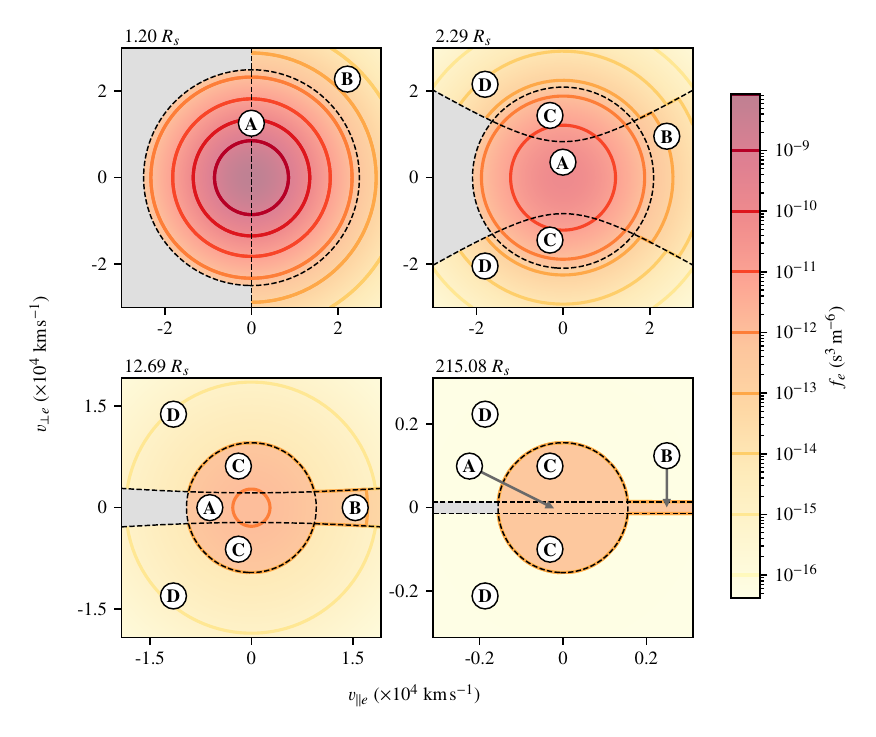}
    \caption{Radial evolution of the electron VDF from the exobase (1.2~$R_s$ in our model) to 1~AU ($\sim$215~$R_s$) modeled by including diffusion for electrons. Region \textbf{D} represents the electron diffusion region where $\zeta = 0.1$ was assumed. }
    \label{fig:eVDFr}
\end{figure*}

Therefore, diffusion is modeled by defining a new parameter that aims to reduce the number of electrons in the \textbf{A}, \textbf{B}, and \textbf{C} regions and adding the same number of electrons to the initially empty region of Fig. \ref{fig:eVDFr} labeled \textbf{D}. This diffusion parameter is defined as
\begin{equation}
    D_r = (1-\eta^{1/12})\zeta,
    \label{eq:D_r}
\end{equation}
so that when $r \rightarrow \infty$ ($\eta \rightarrow 0$) then $D_r \rightarrow \zeta$, with $\zeta$ being a value between 0 and 1 that modulates the relative number of diffused particles, compared to the others (of regions \textbf{A}, \textbf{B}, and \textbf{C}), at an infinite distance from the Sun. The choice that the diffusion profile evolves as $1-(r_0/r)^{6}$ is entirely arbitrary. It was selected to ensure that $D_r = 0$ at $r = r_0$ while also preventing excessively rapid growth, thereby avoiding the diffusion coefficient reaching near its maximum value well before approaching 1~AU.

Furthermore, the density within the combined velocity space region of \textbf{A}, \textbf{B}, and \textbf{C} can be deduced by the addition of Eqs.~(5), (25), and (30) of \cite{pierrard_lorentzian_1996}. This can be written as 
\begin{equation}
    \begin{split}
    n_e(r) = & N_{e} \left(1 + \frac{\psi_e(r)}{\kappa}\right)^{- \kappa - 1} \\
    & \times \paren{\left(1 + \frac{\psi_e(r)}{\kappa}\right)^{3/2} \left(1 - \frac{1}{2} \beta_2(b)\right)
    - \frac{1}{2}\alpha c^{3/2}\beta_2(f)},
    \end{split}
    \label{eq:ne_r}
\end{equation}
with
\begin{equation}
    b = \frac{\kappa + \psi_e(r)}{\kappa + \psi_e(r_\infty)},
\end{equation}
\begin{equation}
    \alpha = \sqrt{1 - \eta} \paren{1 + \frac{\eta}{1-\eta}\left(\frac{\psi_e(r)}{\kappa + \psi_e(r)}\right)}^{-\kappa-1},
\end{equation}
\begin{equation}
    c = 1 + \frac{\psi_e(r)}{(1-\eta)\kappa},
\end{equation}
\begin{equation}
    f = \frac{(1-\eta)\kappa + \psi_e(r)}{(1-\eta)(\kappa + \psi_e(r_\infty))},
\end{equation}
\begin{equation}
    \beta_2(x) = \int_{0}^{x} A_k t^{\kappa - 3/2}(1 - t)^{1/2}\mathrm{d}t,
\end{equation}
\begin{equation}
    A_k = \frac{\Gamma(\kappa + 1)}{\Gamma(\kappa - 1/2)\Gamma(3/2)},
\end{equation}
and 
\begin{equation}
    N_e = \frac{n_0}{1-(1/2)\beta_2(b)},
\end{equation}
such that the density of electrons $n_0$ is equal to the total density of electrons at the exobase since
\begin{equation}
    n_0 \equiv n_e(r_0) = N_e \left(1-\frac{1}{2}\beta_2(b)\right).
\end{equation}
Using the total density and the diffusion parameter, we can now deduce the density of the depleted region (combined region of \textbf{A}, \textbf{B}, and \textbf{C}), which corresponds to
\begin{equation}
    n_e^{\prime}(r) = n_e(r) (1-D_r).
    \label{eq:ne_r_t}
\end{equation}
This implies that the electron density of the new velocity space region \textbf{D} is given by
\begin{equation}
    n_{e,\textbf{D}}(r) = n_e(r) D_r,
    \label{eq:ne_r_i}
\end{equation}
such that the total density remains unchanged since $n_e(r) = n_e^{\prime}(r) + n_{e,\textbf{D}}(r)$. 

The electron flux is here defined using Eq.~(26) of \cite{pierrard_lorentzian_1996} by taking into account the diffusion. This corresponds to the only nonzero contribution of the flux associated with the escaping particles (region \textbf{B}) and written as
\begin{equation}
    F_e^\prime(r) =  \frac{\eta}{4} N_e \frac{w_e}{\sqrt{\kappa}} \frac{A_k}{\kappa - 1} (1 + \psi_e(r_\infty))\left(1 + \frac{\psi_e(r_\infty)}{\kappa}\right)^{-\kappa} (1-D_r).
\end{equation}
From this, it follows that the total flux is $F_e^\prime(r) + F_{e,\textbf{D}}(r) = F_e(r) (1-D_r)$, with $F_{e,\textbf{D}}(r) = 0$ being the diffusion contribution to the flux and $F_e(r)$ being the total flux as if there were no diffusion. Here, the diffused electrons (from the depleted region) are scattered so that they do not contribute to the flux ($F_{e,\textbf{D}}(r) = 0$). This provides a simple approach for formulating the effect of diffusion on the flux in our kinetic exospheric model. We note that the proton flux must be modified accordingly to maintain the zero-current condition that was used to find the solution for the electric potential.

To deduce the electron pressures of the \textbf{D} region, we first defined the pressure without any diffusion of electrons by adding Eqs.~(7), (27), and (31) of \cite{pierrard_lorentzian_1996} for the parallel component to obtain
\begin{equation}
    \begin{split}
    P_{\parallel e} (r) = & \frac{1}{3} N_e m_e w_e^2 \frac{A_k}{A_k^\prime} \kappa \left(1+\frac{\psi_e(r)}{\kappa}\right)^{-\kappa-1} \\
    & \times \Biggl(\left(1 + \frac{\psi_e(r)}{\kappa}\right)^{5/2}\left(1-\frac{1}{2} \beta_4(b)\right)  - \frac{1}{2}\alpha c^{5/2}\left(1-\eta\right)\beta_4(f)\Biggr),
    \end{split}
\end{equation}
and adding Eqs.~(8), (28), and (32) for the perpendicular component to obtain
\begin{equation}
    \begin{split}
        P_{\perp e}(r) = & \frac{1}{3} N_e m_e w_e^2 \frac{A_k}{A_k^\prime} \kappa \left(1+\frac{\psi_e(r)}{\kappa}\right)^{-\kappa-1} \\
        & \times \Biggl(\left(1 + \frac{\psi_e(r)}{\kappa}\right)^{5/2}\left(1- \frac{1}{2} \beta_4(b)\right) - \frac{1}{2}\alpha c^{5/2}\left(1-\frac{\eta}{2}\right)\beta_4(f) \\
        & - \frac{3}{4} \frac{\eta}{(1-\eta)}\frac{\psi_e(r)}{\kappa}\alpha c^{5/2}\beta_2(f) \frac{A_k^\prime}{A_k}\Biggr),
    \end{split}
\end{equation}
with 
\begin{equation}
    A_k^\prime = \frac{\Gamma(\kappa + 1)}{\Gamma(\kappa - 3/2)\Gamma(5/2)}
\end{equation}
and
\begin{equation}
    \beta_4(x) = \int_{0}^{x} A_k^\prime t^{\kappa - 5/2}(1 - t)^{3/2}\mathrm{d}t.
\end{equation}
The parallel and perpendicular pressures for the depleted region are therefore written as
\begin{equation}
    P_{\parallel e}^\prime(r) = P_{\parallel e}(r) (1-D_r)
\end{equation}
and
\begin{equation}
    P_{\perp e}^\prime(r) = P_{\perp e}(r) (1-D_r).
\end{equation}
To determine the pressures in the diffusion region, we introduced the electron density $n_{e,\text{I}}$, obtained by integrating the electron VDF from Eq.~\eqref{eq:fe} within the boundaries of the \textbf{D} region. This quantity corresponds to subtracting Eqs.~(35) with (25) of \cite{pierrard_lorentzian_1996}, as expressed by
\begin{equation}
    n_{e,\text{I}}(r) \equiv N_e \left(1 + \frac{\psi_e(r)}{\kappa}\right)^{- \kappa - 1} \alpha c^{3/2}\beta_2(f)
    \label{eq:ne_r_t_tot},
\end{equation}
which is due to the absence of the incoming particles that would populate the gray region in Fig. \ref{fig:eVDFr}. This gray region is the exospheric equivalent to the observed sunward deficit of electrons in solar wind observations.

In the same way, we defined the parallel and perpendicular pressure by subtracting Eq. (27) from Eq. (37) and Eq. (28) from Eq.~(38), respectively, and multiplying them by the ratio $n_{e,\textbf{D}}(r)/n_{e,\text{I}}(r)$ to obtain
\begin{equation}
    \begin{split}
    P_{\parallel e,\textbf{D}}(r) = & \frac{1}{3} N_{e,\textbf{D}} m_e w_e^2 \frac{A_k}{A_k^\prime} \kappa \left(1+\frac{\psi_e(r)}{\kappa}\right)^{-\kappa-1} \alpha c^{5/2}\left(1-\eta\right)\beta_4(f) 
    \end{split}
\end{equation}
and
\begin{equation}
    \begin{split}
        P_{\perp e,\textbf{D}}(r) = & \frac{1}{3} N_{e,\textbf{D}} m_e w_e^2 \frac{A_k}{A_k^\prime} \kappa \left(1+\frac{\psi_e(r)}{\kappa}\right)^{-\kappa-1} \\
        & \times \Biggl(\alpha c^{5/2}\left(1-\frac{\eta}{2}\right)\beta_4(f) \\
        & + \frac{3}{2} \frac{\eta}{(1-\eta)}\frac{\psi_e(r)}{\kappa}\alpha c^{5/2}\beta_2(f) \frac{A_k^\prime}{A_k}\Biggr).
    \end{split}
\end{equation}
The pressures are those corresponding to the \textbf{D} region of the electron VDF in Fig.~\ref{fig:eVDFr}, where $N_{e,\textbf{D}}$ is the normalization factor of the electron VDF (replacing $N_e$ in Eq.~\eqref{eq:fe}) so that the density of this region is always equal to $n_{e,\textbf{D}}(r)$ as described by Eq.~\eqref{eq:ne_r_i}. This is inferred by noticing that 
\begin{equation}
    \frac{n_{e,\text{I}}(r)}{N_e} = \frac{n_{e,\textbf{D}}(r)}{N_{e,\textbf{D}}}.
\end{equation}
Therefore, we can deduce that $N_{e,\textbf{D}} = N_e n_{e,\textbf{D}}(r)/n_{e,\text{I}}(r)$. 

Following the discussions in the preceding paragraphs, the macroscopic quantities computed in the model for the electrons are the density
\begin{equation}
    n_e^{\prime \prime}(r) \equiv n_e^{\prime}(r) + n_{e,\textbf{D}}(r) = n_e(r),
\end{equation}
the flux
\begin{equation}
    F_e^{\prime \prime}(r) \equiv F_e^\prime(r) + F_{e,\textbf{D}}(r) = F_e^\prime(r),
\end{equation}
the parallel pressure
\begin{equation}
    P_{\parallel e}^{\prime \prime}(r) \equiv P_{\parallel e}^\prime(r) + P_{\parallel e,\textbf{D}}(r),
\end{equation}
and the perpendicular pressure
\begin{equation}
    P_{\perp e}^{\prime \prime}(r) \equiv P_{\perp e}^\prime(r) + P_{\perp e,\textbf{D}}(r).
\end{equation}
This allowed us to then deduce the bulk velocity, 
\begin{equation}
    v_{\text{sw}}(r) \equiv \frac{F_e^{\prime \prime}(r)}{n_e^{\prime \prime}(r)}
,\end{equation}
and total temperatures,
\begin{equation}
    T_{\parallel e} (r) \equiv \frac{1}{k_b}\left(\frac{P_{\parallel e}^{\prime \prime}}{n_e^{\prime \prime}(r)} - m_e v_{\text{sw}}^2\right),
\end{equation}
\begin{equation}
    T_{\perp e} (r) \equiv \frac{1}{k_b}\frac{P_{\perp e}^{\prime \prime}}{n_e^{\prime \prime}(r)}.
\end{equation}
With this relatively simple approach, the influence of diffusion on the radial evolution of the density, bulk velocity, and temperature can be calculated with our kinetic exospheric model. 

\section{Diffusive kinetic exospheric model: Results}

\label{seq:res}

\begin{figure*}[!t]
    \centering
    \begin{tabular}{cc}

        \includegraphics{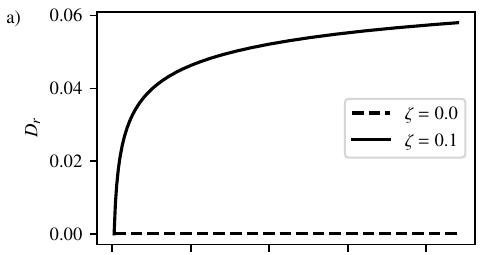} &
        \includegraphics{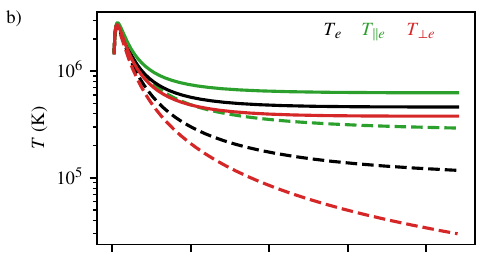} \\
        
        \includegraphics{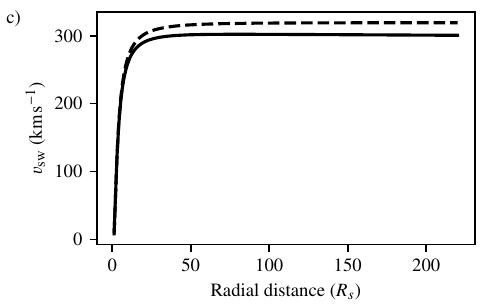} & 
        \includegraphics{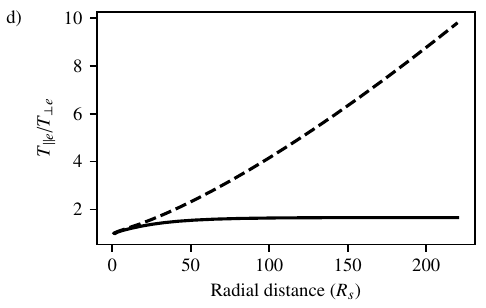}

    \end{tabular}
    \caption{Radial evolution of different quantities obtained by the model with (solid line) or without (dashed line) electron diffusion. Panel a: Electron diffusion parameter. Panel b: Parallel (green), perpendicular (red), and total (black) temperature with respect to the direction of magnetic field lines. Panel c: Bulk velocity. Panel d: Temperature anisotropy ($T_{\parallel e}/T_{\perp e}$).}
    \label{fig:moments}
\end{figure*}

To emphasize the impact of diffusion, we analyzed two scenarios: one that incorporates diffusion and another that does not. We adopted the same fundamental parameters of the exospheric model as those used for the illustrations in Figs.~\ref{fig:VDFcomparison} and \ref{fig:eVDFr}, as specified in Table~\ref{tab:inputs}, and we assumed the diffusion radial evolution described by Eq.~\eqref{eq:D_r} with $\zeta = 0.1$ as the relative maximum amount of diffused particles. This $\zeta$ value was arbitrarily chosen to qualitatively analyze the effect of diffusion, as further improvements to the exospheric diffusion framework would be required to accurately constrain the radial evolution of diffused particles with observations. The radial evolution of the diffusion, electron temperatures, solar wind speed, and electron temperature anisotropy given by our kinetic exospheric model is shown in Fig.~\ref{fig:moments}.

\begin{figure}[!t]
    \centering
    \includegraphics{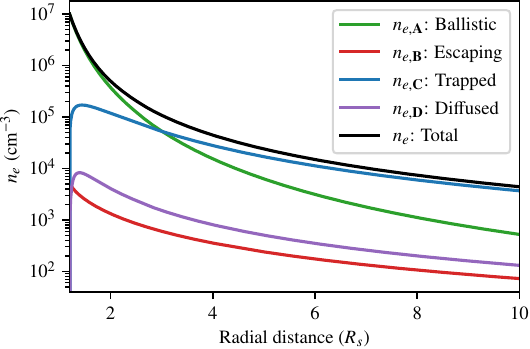}
    \caption{Radial evolution of the different electron population densities computed with the parameters from Table~\ref{tab:inputs} and $\zeta = 0.1$ in our modified kinetic exospheric model.}
    \label{fig:den_e}
\end{figure}

In Fig.~\ref{fig:moments}a, the proportion of diffused particles increases with the radial distance, as expected from Eq.~\eqref{eq:D_r}. The radial evolution of the electron temperatures, when affected by diffusion, shows a slower decrease in Fig.~\ref{fig:moments}b. This behavior is explained by the fact that the electron VDF is essentially broadened by placing particles in the diffusion region. Fig.~\ref{fig:moments}d shows a clear stabilization of the temperature anisotropy, which reaches a value of 1.65 at 215~$R_s$ when diffusion is considered. At 1~AU, observations show that the temperature anisotropy $T_\parallel/T_\perp$ is typically around 1 (that is, without anisotropy) for a slow wind of 300~$\si{km.s^{-1}}$ and increases with the solar wind speed to 1.4 at 700~$\si{km.s^{-1}}$ (\citealt{salem_precision_2023}). In this respect, the naive introduction of diffusion in our model already brings the temperature anisotropy to a value fairly close to the observations in comparison to what was previously found with exospheric models.  

The solar wind speed in Fig.~\ref{fig:moments}c shows a slight decrease when diffusion is included. This is because the diffused particles are assumed to be redistributed without any preferential parallel direction. Therefore, these particles do not contribute to the bulk velocity but, being the only contribution to the flux of the solar wind in our model, do effectively diminish the number of particles within the strahl (essentially the \textbf{B} region). This in turn reduces the bulk velocity shown in Fig.~\ref{fig:moments}c. As a consequence, because of the zero-current condition, the bulk velocity of protons must decrease by the same magnitude as that of the electrons.

The radial evolution of the electron densities is shown in Fig.~\ref{fig:den_e} for the parameters of Table~\ref{tab:inputs} and the same diffusion as previously considered ($\zeta = 0.1$). As mentioned in the previous section, the total density is unaffected by the diffusion. It can be seen that the amount of diffused electrons experiences a rapid growth with radial distances, exceeding the number of escaping electrons (i.e., within the \textbf{B} region) just above the exobase (at 1.2~$R_s$). The diffused particles even exceed the ballistic ones at around 32~$R_s$ (not shown in this figure). This occurs because the trapped particles (that is, those within the \textbf{C} region) rapidly dominate the total electron density. Since the number of diffused particles is determined by the total density, the diffused electrons subsequently outweigh the ballistic ones at large distances.

\section{Exospheric perspective on solar wind acceleration}

\label{seq:dis}

\begin{table*}[!t]
    \caption{\label{tab:inputs_E_pot} Input parameters for four different kinetic exospheric model calculations to represent the electric potentials of Fig.~\ref{fig:Epot}.} 
    \begin{center}
        \begin{tabular}{c c c c c c c}
        Parameter name & Notation & SSW1 & FSW1 & SSW2 & FSW2 & Unit \\
        \hline
        Exobase level & $r_0$ & 1.2 & 1.2 & 7 & 1.01 & $R_s$  \\
        Electron temperature at exobase & $T_{0e}$ & 1.5 & 1.5 & 1 & 1 & $\si{MK}$ \\
        Proton temperature at exobase & $T_{0p}$ & 1.2 & 1.2 & 0.8 & 2 & $\si{MK}$ \\
        Kappa & $\kappa$ & 6 & 3 & 6 & 3 & / \\
        \end{tabular}
    \end{center}
\end{table*}

As presented in Sect.~\ref{seq:exo}, a key advantage of the kinetic exospheric model is its ability to self-consistently determine the electric field under the constraints of quasi-neutrality and zero-current conditions. Consequently, the ambipolar electric field arises directly from the greater influence of the gravitational field on protons compared to that of electrons. This resulting global electric potential contributes, at least partially, to the acceleration of the solar wind and is generally entirely responsible for the acceleration in the context of kinetic exospheric models (\citealt{bercic_interplay_2021,bercic_ambipolar_2021}).

\begin{figure}[!t]
    \centering
    \includegraphics{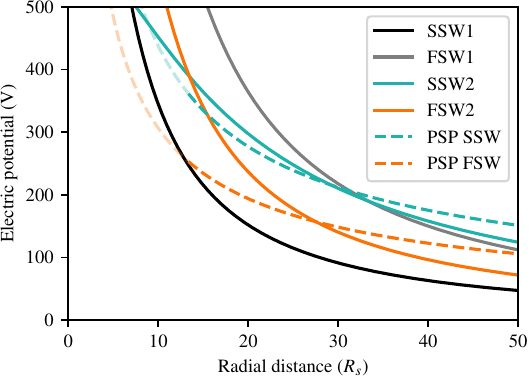}
    \caption{Radial evolution of the electric potential with respect to infinity. The solid lines are the electric potential retrieved by the exospheric model with the input parameters from Table~\ref{tab:inputs_E_pot}. The corresponding solar wind terminal velocities of our model are approximately: 320~$\si{km.s^{-1}}$ for SSW1, 280~$\si{km.s^{-1}}$ for SSW2, 640~$\si{km.s^{-1}}$ for FSW1, and 520~$\si{km.s^{-1}}$ for FSW2. The dashed lines correspond to the SSW (blue) and FSW (orange) limits fitted to PSP data between 13.3 and 50~$R_s$ (encounters 3 to 11) by \cite{halekas_radial_2022}. The dashed blue line is the fit of solar wind data with asymptotic speeds of 200--250~$\si{km.s^{-1}}$, while the dashed orange line corresponds to asymptotic speeds of 600--650~$\si{km.s^{-1}}$.}
    \label{fig:Epot}
\end{figure}

Following the kinetic exospheric approach, a signature of the ambipolar electric potential can be identified in the solar wind data. Specifically, it can be inferred from the cutoff velocity associated with the deficit in the sunward direction by analyzing the observed electron VDFs in the solar wind (\citealt{halekas_electrons_2020,bercic_ambipolar_2021,bercic_whistler_2021,halekas_sunward_2021,halekas_radial_2022}). Introducing the asymptotic wind speed, \cite{halekas_radial_2022} binned the cutoff velocities of PSP between 13.3 and 50~$R_s$ (for encounters 3 to 11). In this way, the radial evolution of the electric potential associated with asymptotic speeds of 200--250 and 600--650~$\si{km.s^{-1}}$ has been deduced and is represented in Fig.~\ref{fig:Epot} for PSP slow solar wind (SSW) and PSP fast solar wind (FSW). An anticorrelation between the electric potential and the velocity of the solar wind can be clearly identified (\citealt{halekas_radial_2022}). 

For the kinetic exospheric model, the radial evolution of the self-consistent electric potential is shown in Fig.~\ref{fig:Epot} for four different sets of parameters given by Table~\ref{tab:inputs_E_pot}. The SSW1 solid black line is a SSW solution found using the same test parameters as in Table~\ref{tab:inputs}, which was introduced to illustrate the effect of diffusion. We note that the electric potential is independent of the diffusion, since it is found by using the quasi-neutrality with an unchanged total density. It is commonly understood that exospheric models predict a faster solar wind accompanied by a larger electric potential at any radial distance, to account for the enhanced acceleration of the solar wind. This contradiction with the observed anticorrelation mentioned above holds true only if we consider that the acceleration is entirely determined by the $\kappa$ parameter. As illustrated in Fig.~\ref{fig:Epot}, this is evident from the comparison between SSW1 and FSW1, where reducing $\kappa$ from 6 to 3 results in a corresponding increase in the terminal wind velocity from 320 to 640~$\si{km.s^{-1}}$. However, when considering the same change in $\kappa$ for SSW2 and FSW2, a carefully selected combination of exobase levels and initial temperatures (see Table~\ref{tab:inputs_E_pot}) produces a crossing point between the two electric potentials (at around 13~$R_s$) while still maintaining the increase in terminal wind velocity from 280 (for SSW2) to 520~$\si{km.s^{-1}}$ (for FSW2). These parameters were chosen because they approximate the observed behavior to some extent, producing a crossing point at low altitude while maintaining realistic values (as described in the next paragraph). However, a wide range of parameter sets can reproduce a crossing point, provided they meet two key conditions: (1) a sufficiently large difference between proton and electron temperatures (compared to the difference in kappa values) in fast wind relative to slow wind, and (2) a significantly higher exobase level for slow wind. In this way, the anticorrelation between the electric potential and the solar wind terminal velocity is retrieved above 13~$R_s$ with the kinetic exospheric model. This indicates that the kinetic exospheric model may be better suited than previously thought for explaining the FSW through the electric field, especially if the crossing point between the electric potentials of fast and slow winds occurs below 13~$R_s$. Although PSP observations do not appear to indicate the presence of such an intersection, or at least not above 13~$R_s$, the consistency of our modeled electric potentials compared with observations suggests that further investigation of the most recent PSP data is needed to completely rule out this possibility. 

This consideration is reinforced by the fact that the parameters found to reproduce the anticorrelation between the electric potential and the solar wind speed (see SSW2 and FSW2 of Table~\ref{tab:inputs_E_pot}) are also consistent with the deduced coronal temperatures (\citealt{david_measurement_1998,cranmer_coronal_2002}). The FSW typically appears above coronal holes that are characterized by a higher proton temperature compared to that of electrons and a low density that lowers the exobase compared to other regions of the Sun's surface (\citealt{pierrard_exospheric_2023}), which is the case for FSW2. This is also supported by the anticorrelation between the strahl parallel temperature and the solar wind speed, consistently found by Helios (65--215~$R_s$) and PSP (13.3--50~$R_s$; \citealt{bercic_coronal_2020,halekas_radial_2022}). Furthermore, as identified with SSW2, the SSW is commonly associated with denser and less active regions of the equatorial corona, where the exobase is therefore higher with a generally hotter electron temperature than the coronal hole conditions (\citealt{david_measurement_1998}). As reported by \citealt{bercic_coronal_2020}, these findings suggest that the strahl parallel temperature preserves the electron coronal temperature. The exospheric approach is consistent with this idea by maintaining the strahl parallel temperature constant throughout the solar wind expansion, since no collisions nor wave-particle interactions are assumed.  

It is important to note that the implementation of the diffusion mechanism does not inherently require that particles in region \textbf{D} do not contribute to the flux. For that reason, an additional acceleration mechanism could be taken into account within the current diffusion framework by considering that the transfer of core particles (combined ballistic and trapped particles of regions \textbf{A} and \textbf{C}) to higher energies contributes to the flux. In practice, this transfer would increase the number of escaping particles and/or enhance the anti-sunward portion of the halo (compared to the sunward portion). Another refinement for better alignment with observations would be to account for diffusion creating a proper sunward deficit, rather than just an empty region, along with its associated negative flux contribution. The key idea is that the sum of all newly introduced flux contributions from diffusion (of the core) could ultimately result in a net increase in total flux, thereby providing an additional acceleration mechanism within the kinetic exospheric model.

One limitation to consider is that the escaping particles (region \textbf{C}) defined within our current kinetic exospheric approach cannot be the sole source of halo electrons. A quantitative analysis of our exospheric model results (neglecting diffusion) indicates that the relative density of escaping electrons, compared to the total density, is approximately 3\% at 65~$R_s$, increasing to about 4\% at 215~$R_s$, with minimal variations across the parameter sets used in this study (SSW1, FSW1, SSW2, and FSW2). However, observations by \cite{stverak_radial_2009} suggest that the combined halo and strahl population should constitute roughly 7.5\% of the total electron density over the same radial distances. This discrepancy implies that escaping particles alone cannot account for the observed halo electron population within our current exospheric model framework.

Taking this into account, we introduced a diffusion that transfers particles from the strahl to the halo, which improves the representation of the combined halo and strahl populations, yielding a relative density of approximately 8\% from 65 to 215~$R_s$, in better agreement with observations. Importantly, the primary objective of this study is to propose a modification to the exospheric framework that incorporates diffusion in the simplest possible manner, as this represents the first exospheric model of its kind. However, existing diffusion models (e.g., \citealt{verscharen_self-induced_2019, jeong_quasi-linear_2020, jeong_kinetic_2022}), supported by observations of the strahl and halo populations (e.g., \citealt{stverak_electron_2008, stverak_radial_2009}), indicate that diffusion primarily occurs from the strahl to the halo. This suggests that further refinement of our kinetic exospheric model is necessary to ensure consistency with established models and observations.

The observed suprathermal electrons, included in our existing kinetic exospheric model, constitute an important property of the strahl as they may have significant implications for the solar wind acceleration. According to \cite{Lazarbook_2021_5}, spectral line analyzes seem to indicate that suprathermal populations are present in the corona. However, in situ data from PSP presented by \cite{halekas_radial_2022} suggest that no significant strahl suprathermal electrons are observed near the Sun. This observation implies that Kappa electron VDFs may not be the most likely explanation for the FSW. However, our results, consistent with the findings of \cite{zouganelis_transonic_2004}, indicate that the key factor influencing the solar wind's terminal velocity is not solely the suprathermal component, but rather the relative abundance of particles within the strahl compared to the deficit in the sunward direction.

\section{Conclusions and future perspectives}

\label{seq:concl}

In this paper we have underlined the fundamentals of the existing kinetic exospheric model, discussing possible future improvements of this approach and providing a basis for investigating the influence of electron diffusion on solar wind acceleration. The improvement explored in this study involves redistributing particles within the traditionally considered regions of velocity space --- namely ballistic, trapped, and escaping particles --- in kinetic exospheric models. This method produces diffused particles capable of reaching perpendicular speeds that would otherwise be inaccessible under strict conservation laws. This constitutes a step toward a model that generates more realistic distribution functions based on the kinetic exospheric theory. In this way, we were able to identify several important features of the kinetic exospheric framework applied to solar wind acceleration: 
\begin{enumerate}
    \item A simple implementation of electron diffusion effectively reduces the temperature anisotropy to values comparable to those of the observed FSW at 1~AU.
    \item This diffusion enables us to adjust the solar wind acceleration without modifying the self-consistent ambipolar electric potential found by the kinetic exospheric model.
    \item The observed anticorrelation between wind speed and terminal velocity can be reproduced using the kinetic exospheric model with realistic temperatures and exobase levels. This suggests that further analysis of PSP data is necessary to either confirm or rule out the electric potential as the primary acceleration mechanism for the FSW.
    \item If no indication of the presence of a crossing point between the deduced electric potentials of the FSW and SSW is found within PSP data, an alternative mechanism --- likely a velocity-space diffusion of particles via wave-particle interactions --- must dominate the FSW acceleration by increasing the flux in the anti-sunward direction.
\end{enumerate}

The extension of the kinetic exospheric model by incorporating an additional acceleration mechanism through diffusion presents a promising approach for improving the model, especially if no evidence of an intersection between the electric potentials of the SSW and FSW is detected in PSP data. In doing so, it would be possible to determine how much diffusion-dependent net flux is required to explain the observed profiles of the electric potential with their respective terminal velocity. This investigation is left for future work, as the primary objective of this paper is to present the concept of diffusion within the kinetic exospheric model and its ability to provide another source of acceleration (or deceleration).

The primary limitation of the current diffusion implementation is that it relies solely on the initial temperature, $T_0$, to determine the slope of the diffused population VDF. As a result, the predicted temperatures are overestimated in comparison with observational data. This limitation could be addressed in the future by constraining the slope on the basis of the observed halo or strahl populations and their radial evolution.

Furthermore, greater consistency with the current understanding of the diffusion mechanism could be achieved by explicitly distinguishing between the core and escaping populations, as our current approach considers that the same electron VDF defines both the core and strahl electrons. This refinement could be implemented following the methodology of \citealt{zouganelis_transonic_2004}, in which a Maxwellian distribution and a Kappa distribution are used to define the core and strahl, respectively.

In our exospheric model, the previously introduced Kappa distribution function effectively simulates an increase in the strahl population, thereby accelerating the solar wind to higher velocities. However, this model does not necessarily require non-Maxwellian distributions at the base of the corona to explain the solar wind acceleration; instead, it relies on a significant strahl population near the Sun. Consequently, the critical question for understanding the solar wind acceleration within the framework of the kinetic exospheric model becomes: How is the strahl formed in the low corona? Further investigations using fully kinetic simulations will be conducted to address this question.

\begin{acknowledgements}
M.P.d.B.\ is supported by the FWO Junior Research Project G020224N granted by the Research Foundation -- Flanders (FWO). 
F.B.\ acknowledges support from the FED-tWIN programme (profile Prf-2020-004, project ``ENERGY'') issued by BELSPO.

\end{acknowledgements}

\bibliographystyle{aa}
\bibliography{aa54250-25corr_arXiv.bib}

\begin{thebibliography}{51}
\expandafter\ifx\csname natexlab\endcsname\relax\def\natexlab#1{#1}\fi

\bibitem[{Abraham {et~al.}(2022)Abraham, Owen, Verscharen, Bakrania, Stansby,
  Wicks, Nicolaou, Whittlesey, Agudelo~Rueda, Jeong, \&
  Berčič}]{abraham_radial_2022}
Abraham, J.~B., Owen, C.~J., Verscharen, D., {et~al.} 2022, ApJ, 931, 118

\bibitem[{Berčič {et~al.}(2021{\natexlab{a}})Berčič, Landi, \&
  Maksimović}]{bercic_interplay_2021}
Berčič, L., Landi, S., \& Maksimović, M. 2021{\natexlab{a}}, J. Geophys.
  Res. Space Phys., 126, e2020JA028864

\bibitem[{Berčič {et~al.}(2020)Berčič, Larson, Whittlesey, Maksimović,
  Badman, Landi, Matteini, Bale, Bonnell, Case, Dudok~de Wit, Goetz, Harvey,
  Kasper, Korreck, Livi, MacDowall, Malaspina, Pulupa, \&
  Stevens}]{bercic_coronal_2020}
Berčič, L., Larson, D., Whittlesey, P., {et~al.} 2020, ApJ, 892, 88

\bibitem[{Berčič {et~al.}(2021{\natexlab{b}})Berčič, Maksimović, Halekas,
  Landi, Owen, Verscharen, Larson, Whittlesey, Badman, Bale, Case, Goetz,
  Harvey, Kasper, Korreck, Livi, MacDowall, Malaspina, Pulupa, \&
  Stevens}]{bercic_ambipolar_2021}
Berčič, L., Maksimović, M., Halekas, J.~S., {et~al.} 2021{\natexlab{b}},
  ApJ, 921, 83

\bibitem[{Berčič {et~al.}(2021{\natexlab{c}})Berčič, Verscharen, Owen,
  Colomban, Kretzschmar, Chust, Maksimovic, Kataria, Anekallu, Behar,
  Berthomier, Bruno, Fortunato, Kelly, Khotyaintsev, Lewis, Livi, Louarn, Mele,
  Nicolaou, Watson, \& Wicks}]{bercic_whistler_2021}
Berčič, L., Verscharen, D., Owen, C.~J., {et~al.} 2021{\natexlab{c}}, A\&A,
  656, A31

\bibitem[{Boldyrev {et~al.}(2020)Boldyrev, Forest, \&
  Egedal}]{boldyrev_electron_2020}
Boldyrev, S., Forest, C., \& Egedal, J. 2020, Proc. Natl. Acad. Sci. USA, 117,
  9232

\bibitem[{Chamberlain(1960)}]{chamberlain_interplanetary_1960}
Chamberlain, J.~W. 1960, ApJ, 131, 47

\bibitem[{Cranmer(2002)}]{cranmer_coronal_2002}
Cranmer, S. 2002, in COSPAR Colloquia Series, Vol.~13, Multi-wavelength
  Observations of Coronal Structure and Dynamics, ed. P.~C. Martens \& D.~P.
  Cauffman (Pergamon), 3--12

\bibitem[{David {et~al.}(1998)David, Gabriel, Bely-Dubau, Fludra, Lemaire, \&
  Wilhelm}]{david_measurement_1998}
David, C., Gabriel, A.~H., Bely-Dubau, F., {et~al.} 1998, A\&A, 336, L90

\bibitem[{Effenberger \& Jeffrey(2021)}]{Lazarbook_2021_5}
Effenberger, F. \& Jeffrey, N. L.~S. 2021, in Kappa Distributions: From
  Observational Evidences via Controversial Predictions to a Consistent Theory
  of Nonequilibrium Plasmas, ed. M.~Lazar \& H.~Fichtner (Cham: Springer
  International Publishing), 89--103

\bibitem[{Fahr \& Shizgal(1983)}]{fahr_modern_1983}
Fahr, H.~J. \& Shizgal, B. 1983, Rev. Geophys., 21, 75

\bibitem[{Feldman {et~al.}(1975)Feldman, Asbridge, Bame, Montgomery, \&
  Gary}]{feldman_solar_1975}
Feldman, W.~C., Asbridge, J.~R., Bame, S.~J., Montgomery, M.~D., \& Gary, S.~P.
  1975, J. Geophys. Res., 80, 4181

\bibitem[{Halekas {et~al.}(2021)Halekas, Berčič, Whittlesey, Larson, Livi,
  Berthomier, Kasper, Case, Stevens, Bale, MacDowall, \&
  Pulupa}]{halekas_sunward_2021}
Halekas, J.~S., Berčič, L., Whittlesey, P., {et~al.} 2021, ApJ, 916, 16

\bibitem[{Halekas {et~al.}(2022)Halekas, Whittlesey, Larson, Maksimovic, Livi,
  Berthomier, Kasper, Case, Stevens, Bale, MacDowall, \&
  Pulupa}]{halekas_radial_2022}
Halekas, J.~S., Whittlesey, P., Larson, D.~E., {et~al.} 2022, ApJ, 936, 53

\bibitem[{Halekas {et~al.}(2020)Halekas, Whittlesey, Larson, McGinnis,
  Maksimovic, Berthomier, Kasper, Case, Korreck, Stevens, Klein, Bale,
  MacDowall, Pulupa, Malaspina, Goetz, \& Harvey}]{halekas_electrons_2020}
Halekas, J.~S., Whittlesey, P., Larson, D.~E., {et~al.} 2020, ApJS, 246, 22

\bibitem[{Hansteen \& Velli(2012)}]{hansteen_solar_2012}
Hansteen, V.~H. \& Velli, M. 2012, Space Sci. Rev., 172, 89

\bibitem[{Hollweg \& Isenberg(2002)}]{hollweg_generation_2002}
Hollweg, J.~V. \& Isenberg, P.~A. 2002, J. Geophys. Res., 107, 1147

\bibitem[{Jeong {et~al.}(2022)Jeong, Verscharen, Vocks, Abraham, Owen, Wicks,
  Fazakerley, Stansby, Berčič, Nicolaou, Agudelo~Rueda, \&
  Bakrania}]{jeong_kinetic_2022}
Jeong, S.-Y., Verscharen, D., Vocks, C., {et~al.} 2022, ApJ, 927, 162

\bibitem[{Jeong {et~al.}(2020)Jeong, Verscharen, Wicks, \&
  Fazakerley}]{jeong_quasi-linear_2020}
Jeong, S.-Y., Verscharen, D., Wicks, R.~T., \& Fazakerley, A.~N. 2020, ApJ,
  902, 128

\bibitem[{Jockers(1970)}]{jockers_solar_1970}
Jockers, K. 1970, A\&A, 6, 219

\bibitem[{Lamy {et~al.}(2003)Lamy, Pierrard, Maksimovic, \&
  Lemaire}]{lamy_kinetic_2003}
Lamy, H., Pierrard, V., Maksimovic, M., \& Lemaire, J.~F. 2003, J. Geophys.
  Res., 108, 1047

\bibitem[{Leer {et~al.}(1982)Leer, Holzer, \& Fl\r{A}}]{leer_acceleration_1982}
Leer, E., Holzer, T.~E., \& Fl\r{A}, T. 1982, Space Sci. Rev., 33, 161

\bibitem[{Lemaire \& Scherer(1970)}]{lemaire_model_1970}
Lemaire, J. \& Scherer, M. 1970, P\&SS, 18, 103

\bibitem[{Lemaire \& Scherer(1971)}]{lemaire_kinetic_1971}
Lemaire, J. \& Scherer, M. 1971, J. Geophys. Res., 76, 7479

\bibitem[{Lin(1980)}]{lin1980energetic}
Lin, R. 1980, Sol. Phys., 67, 393

\bibitem[{Maksimovic {et~al.}(1997{\natexlab{a}})Maksimovic, Pierrard, \&
  Lemaire}]{maksimovic_kinetic_1997}
Maksimovic, M., Pierrard, V., \& Lemaire, J.~F. 1997{\natexlab{a}}, A\&A, 324,
  725

\bibitem[{Maksimovic {et~al.}(1997{\natexlab{b}})Maksimovic, Pierrard, \&
  Riley}]{maksimovic_ulysses_1997}
Maksimovic, M., Pierrard, V., \& Riley, P. 1997{\natexlab{b}}, Geophys. Res.
  Lett., 24, 1151

\bibitem[{Micera {et~al.}(2020)Micera, Zhukov, López, Innocenti, Lazar,
  Boella, \& Lapenta}]{micera_particle--cell_2020}
Micera, A., Zhukov, A.~N., López, R.~A., {et~al.} 2020, ApJL, 903, L23

\bibitem[{Parker(1965)}]{parker_dynamical_1965}
Parker, E. 1965, Space Sci. Rev., 4, 666

\bibitem[{Pierrard {et~al.}(2001{\natexlab{a}})Pierrard, Issautier,
  Meyer‐Vernet, \& Lemaire}]{pierrard_collisionless_2001}
Pierrard, V., Issautier, K., Meyer‐Vernet, N., \& Lemaire, J.
  2001{\natexlab{a}}, Geophys. Res. Lett., 28, 223

\bibitem[{Pierrard \& Lazar(2010)}]{pierrard_kappa_2010}
Pierrard, V. \& Lazar, M. 2010, Sol. Phys., 267, 153

\bibitem[{Pierrard {et~al.}(2016)Pierrard, Lazar, Poedts, Stverak, Maksimovic,
  \& Travnicek}]{pierrard_halo2016}
Pierrard, V., Lazar, M., Poedts, S., {et~al.} 2016, J. Geophys. Res. Space
  Phys., 291, 2165

\bibitem[{Pierrard {et~al.}(2011)Pierrard, Lazar, \&
  Schlickeiser}]{pierrard_evolution_2011}
Pierrard, V., Lazar, M., \& Schlickeiser, R. 2011, Sol. Phys., 269, 421

\bibitem[{Pierrard \& Lemaire(1996)}]{pierrard_lorentzian_1996}
Pierrard, V. \& Lemaire, J. 1996, J. Geophys. Res. Space Phys., 101, 7923

\bibitem[{Pierrard {et~al.}(2001{\natexlab{b}})Pierrard, Maksimovic, \&
  Lemaire}]{pierrard_strahl2001}
Pierrard, V., Maksimovic, M., \& Lemaire, J. 2001{\natexlab{b}}, Astrophys.
  Space Sciences, 277, 195

\bibitem[{Pierrard {et~al.}(2001{\natexlab{c}})Pierrard, Maksimovic, \&
  Lemaire}]{pierrard_selfconsistent_2001}
Pierrard, V., Maksimovic, M., \& Lemaire, J. 2001{\natexlab{c}}, J. Geophys.
  Res. Space Phys., 106, 29305

\bibitem[{Pierrard \& Pieters(2014)}]{pierrard_coronal_2014}
Pierrard, V. \& Pieters, M. 2014, J. Geophys. Res. Space Phys., 119, 9441

\bibitem[{Pierrard {et~al.}(2023)Pierrard, Péters~de Bonhome, Halekas, Audoor,
  Whittlesey, \& Livi}]{pierrard_exospheric_2023}
Pierrard, V., Péters~de Bonhome, M., Halekas, J., {et~al.} 2023, Plasma, 6,
  518

\bibitem[{Pilipp {et~al.}(1987)Pilipp, Miggenrieder, Montgomery, Mühlhäuser,
  Rosenbauer, \& Schwenn}]{pilipp_characteristics_1987}
Pilipp, W.~G., Miggenrieder, H., Montgomery, M.~D., {et~al.} 1987, J. Geophys.
  Res. Space Phys., 92, 1075

\bibitem[{Rouillard {et~al.}(2021)Rouillard, Viall, Pierrard, Vocks, Matteini,
  Alexandrova, Higginson, Lavraud, Lavarra, Wu, Pinto, Bemporad, \&
  Sanchez‐Diaz}]{raouafi_solar_2021}
Rouillard, A.~P., Viall, N., Pierrard, V., {et~al.} 2021, in Geophysical
  {Monograph} {Series}, 1st edn., ed. N.~E. Raouafi, A.~Vourlidas, Y.~Zhang, \&
  L.~J. Paxton (Wiley), 1--33

\bibitem[{Salem {et~al.}(2003)Salem, Hubert, Lacombe, Bale, Mangeney, Larson,
  \& Lin}]{salem_electron_2003}
Salem, C., Hubert, D., Lacombe, C., {et~al.} 2003, ApJ, 585, 1147

\bibitem[{Salem {et~al.}(2023)Salem, Pulupa, Bale, \&
  Verscharen}]{salem_precision_2023}
Salem, C.~S., Pulupa, M., Bale, S.~D., \& Verscharen, D. 2023, A\&A, 675, A162

\bibitem[{Seough {et~al.}(2015)Seough, Nariyuki, Yoon, \&
  Saito}]{seough_strahl_2015}
Seough, J., Nariyuki, Y., Yoon, P.~H., \& Saito, S. 2015, ApJ, 811, L7

\bibitem[{{\v{S}}tverák {et~al.}(2009){\v{S}}tverák, Maksimovic,
  Trávníček, Marsch, Fazakerley, \& Scime}]{stverak_radial_2009}
{\v{S}}tverák, {\v{S}}., Maksimovic, M., Trávníček, P.~M., {et~al.} 2009,
  J. Geophys. Res. Space Phys., 114, A05104

\bibitem[{{\v{S}}tverák {et~al.}(2008){\v{S}}tverák, Trávníček,
  Maksimovic, Marsch, Fazakerley, \& Scime}]{stverak_electron_2008}
{\v{S}}tverák, {\v{S}}., Trávníček, P., Maksimovic, M., {et~al.} 2008, J.
  Geophys. Res. Space Phys., 113, A03103

\bibitem[{Verscharen {et~al.}(2022)Verscharen, Chandran, Boella, Halekas,
  Innocenti, Jagarlamudi, Micera, Pierrard, Štverák, Vasko, Velli, \&
  Whittlesey}]{verscharen_electron-driven_2022}
Verscharen, D., Chandran, B. D.~G., Boella, E., {et~al.} 2022, Front. Astron.
  Space Sci., 9, 951628

\bibitem[{Verscharen {et~al.}(2019)Verscharen, Chandran, Jeong, Salem, Pulupa,
  \& Bale}]{verscharen_self-induced_2019}
Verscharen, D., Chandran, B. D.~G., Jeong, S.-Y., {et~al.} 2019, ApJ, 886, 136

\bibitem[{Vocks {et~al.}(2005)Vocks, Salem, Lin, \& Mann}]{vocks_electron_2005}
Vocks, C., Salem, C., Lin, R.~P., \& Mann, G. 2005, ApJ, 627, 540

\bibitem[{Yoon {et~al.}(2024)Yoon, Salem, Klein, Martinović, López, Seough,
  Sarfraz, Lazar, \& Shaaban}]{yoon_regulation_2024}
Yoon, P.~H., Salem, C.~S., Klein, K.~G., {et~al.} 2024, ApJ, 975, 105

\bibitem[{Zheng {et~al.}(2024)Zheng, Martinović, Liu, Pierrard, Liu, \&
  Cheng}]{Zheng_kappa2024}
Zheng, X., Martinović, M., Liu, K., {et~al.} 2024, ApJ, 977, 39

\bibitem[{Zouganelis {et~al.}(2004)Zouganelis, Maksimovic, Meyer‐Vernet,
  Lamy, \& Issautier}]{zouganelis_transonic_2004}
Zouganelis, I., Maksimovic, M., Meyer‐Vernet, N., Lamy, H., \& Issautier, K.
  2004, ApJ, 606, 542

\end{thebibliography}

\end{document}